\documentstyle[aps,prl,twocolumn]{revtex}

\input epsf 
\draft
\begin{document} 
\title {Integrable Structure of Interface Dynamics\\} 
\author {Mark Mineev-Weinstein} 
\address{Theoretical Division, MS-B213, LANL, 
Los Alamos, NM 87545, USA } 
\author{Paul B. Wiegmann} 
\address{James Franck Institute and Enrico Fermi Institute of the 
University of Chicago \\ 
5640 S.Ellis Avenue, Chicago, IL 60637, USA and Landau Institute for 
Theoretical Physics} 
\author{Anton Zabrodin} 
\address{Joint Institute of Chemical Physics, Kosygina str. 4, 
117334, Moscow, Russia \\ 
and ITEP, 117259, Moscow, Russia} 
 
\maketitle 
 
\begin{abstract} 
We establish the equivalence of 2D contour dynamics 
to the dispersionless limit of the integrable Toda hierarchy 
constrained by a string equation. Remarkably, the same hierarchy 
underlies 2D quantum gravity. 
\end{abstract} 
\pacs{PACS number(s):02.30.Em, 02.30,.Dk, 68.10.-m, 68.70.+w, 04.60.-m} 
 
\vspace{.2in} 
 
{\it 1.} {\it Laplacian growth}. Contour dynamics takes place in 
many physical processes, where an interface moves between two immiscible 
phases. The key example of interface dynamics to illustrate the
main result of this work 
is the Laplacian growth  (LG) \cite{1}. This process is dissipative, 
unstable, ubiquitous (applications range from  oil/gas recovery to tumor
growth), and universal: a steady self-similar 
pattern appears governed by scaling laws, most of which still yet to be 
derived \cite {1,2}. 
 
In this paper we show that an arbitrary interface dynamics has an integrable 
structure which is the same as the one that underlies models 
of 2D quantum gravity. 
This structure links the 
interface dynamics, and especially LG, 
with other branches of theoretical physics, where 
scaling laws are also expected 
\cite{GM}. 
 
{\it 2.} To be specific, we will speak about Hele-Shaw flow \cite{1}: 
a viscous fluid (oil) and a non-viscous fluid (water) are confined in a narrow gap 
between two parallel plates.  The interior water domain, $D_+$, is surrounded 
by an exterior oil domain, $D_-$, occupying the rest of the plane. 
Water is supplied from the origin and pushes the oil/water interface, 
${\cal C}(t)$. Both liquids are incompressible, so oil is extracted at infinity 
at the same rate $q$ as water is supplied. 
 
\begin{itemize} 
\item[--] 
\noindent the normal velocity of the interface is 
$V_n =-\partial_n p$ (the D'Arcy law); the pressure $p$ is kept constant ($p=0$) 
inside the water domain $D_+(t)$ and on the interface (surface tension and 
viscosity of the water are neglected); and pressure is a harmonic function, 
$\nabla^2 p = 0$, inside the oil domain $D_-(t)$, while $p \to -(q/2\pi)\log 
\sqrt{x^2+y^2}$ at infinity. 
\end{itemize} 
This (idealized) LG problem has an important property: The harmonic 
moments of the oil domain 
$C_k = \int_{D_-(t)}z^{-k} dx\,dy$ ($k=1, 2,\ldots$, and $z=x+iy$)\footnote{$C_1$ and $C_2$ are finite: the divergence
as $|z|\to \infty$ cancels by integration over $\mbox{arg}\,z$.}  do not 
change in time, while the area of water domain, grows linearly in time 
\cite{Richardson}. The proof: 
$$\frac{d C_k}{d t} = \oint_{{\cal C} (t)} \frac{V_n\,d{\cal C}}{z^k} 
= \oint_{{\cal C} 
(t)}(p\,\partial_n z^{-k} - z^{-k}\partial_n p)\,d\cal C,$$ because 
$V_n = -\partial_n p$ 
and $p=0$ along the ${\cal C}(t)$ . By virtue of the Gauss' 
theorem, it equals 
$$  =\int_{D_-(t)} \nabla\, (p\,\nabla z^{-k} - z^{-k}\nabla p)dx\,dy = 
-q\delta_{k,0}.$$ 
This property may be used as the definition of the idealized LG problem: 
 
\begin{itemize} 
\item[--] 
To find the form of the domain whose area increases 
while all harmonic moments remain fixed. 
\end{itemize} 
 
This problem is known to be ill-defined \cite{1}. For almost 
all sets of harmonic moments, the boundary develops cusp-like 
singularities in finite time (area) \cite{1}. 
Once a singularity occurs, the idealized LG model is no longer valid. 
Surface tension, omitted above, stabilizes the growth 
and simultaneously ruins 
the conservation of harmonic moments. Simulations and experiments show that 
different  mechanisms of regularization of singularities (surface tension, 
lattice etc.) exhibit the same self-similar pattern \cite{1,2}. This 
suggests a fixed point (or points) in the space of harmonic moments, which 
correspond to observed stable patterns. To identify the fixed points and 
their scaling properties is the challenge of the growth phenomena. 
 
Approach to a fixed point requires a 
change of {\it all} moments. 
This is the question we address in this paper. 
 
We 
present the set of differential equations which describe the 
evolution of the domain under a variation of {\it all} 
harmonic moments. 
This prompts to a connection with the inverse 
potential problem \cite{inverse}: to restore the shape of a body 
from a given  Newtonian potential of a uniform mass distribution 
inside the body.

It remains to be seen whether these equations help to 
describe the pattern of growth; however, they reveal the integrable structure 
of the growth problems. We will show that the equations describing the 
evolution of a domain form an integrable hierarchy. Moreover, the very same 
hierarchy emerges in $c=1$ string theory and topological gravity \cite{c=1}, 
and in 2-matrix models \cite{2matrix}. 
It is the dispersionless limit of the 2D 
Toda hierarchy \cite{dToda} constrained by the so-called string equation 
\cite{string}. 
 
{\it 3.} To proceed further we need some known 
facts about the Schwarz function. (See e.g.,\,\cite{Davies}). 
An equation for a curve, 
$F_{\cal C}(x,y) = F_{\cal C}(\frac{z+\bar z}{2}, \frac{z-\bar z}{2i})= 0$, 
can be resolved (at least locally) with respect to one of the complex 
variables, say $\bar z=x-iy$. The result, 
$\bar z = S(z)$, 
is called the {\it Schwarz function} of the curve $\cal C$ (see e.g., 
\cite{Davies}): 
\newline (a) $S(z)$ is a unitary operation: 
$\bar S (S(z) ) = z $; 
\newline (b) The unit  vector tangential to the curve is 
$dz /dl = 
dz/\sqrt{dz d \bar z} = 
(d\bar z(z)/d z)^{-1/2}= 
1/\sqrt {S_z} = 
\sqrt {\bar S_{\bar z}}$; 
\newline (c) For 
simple analytic curves, the Schwarz function can be 
analytically continued to some strip-like domain containing 
the curve. 
 
The function $S(z)$ can be decomposed into a sum of two functions 
$S^{(\pm )}(z)$ that are regular in 
$D_{\pm}$: $S(z) =S^{(+)}(z) + S^{(-)}(z)$. 
Under the condition $S^{(-)}(\infty )=0$ this decomposition is unique. The 
functions $S^{(\pm )}(z)$ can be 
represented by a Taylor series convergent near 
the origin (which is assumed to be 
in $D_+$) and near infinity in $D_-$: 
$S^{(+)}(z)=\sum_{k=0}^{\infty}S_k z^k$, 
$S^{(-)}(z)=\sum_{k=1}^{\infty}S_{-k}z^{-k}$, 
The coefficients $S_{\pm k}$ are nothing but 
harmonic moments of the exterior, $D_-$, and the interior, 
$D_+$, domains: 
\begin{equation} 
C_{\pm k} \!=\! \mp \int_{D_{\mp}} \!\! z^{\mp k}dxdy  \!=\! 
\oint_{{\cal C}(t)} \! \frac {S(z)\,dz}{2iz^{\pm k}} \!=\! 
\pi S_{{\pm}k-1}\,. 
\end{equation} 
In other words, 
$S^{(\pm )}(z)$ is the gradient of  the Newtonian potential 
created by matter uniformly distributed in the 
interior (exterior) of $\cal C$. 
In these terms  the idealized LG problem 
implies that $S^{(+)}$ does not vary in time and $\pi S_{-1}=(\mbox{Area 
of}\; D_+$) grows linearly in time. 
 
The Schwarz function is closely related to conformal maps. 
Let $\phi(x,y)$ be the function harmonically conjugate 
to $2\pi p(x,y)/q$. Then $w = e^{-2\pi p/q + i\phi}$ 
univalently maps the oil domain to the exterior of the unit circle. 
This map sends $w=\infty$ to $z=\infty$. 
Let us write 
\begin{equation} 
\label{z1} 
z(t,w) = r(t)\,w +\sum_{k=0}^{\infty} u_k(t) \,w^{-k} 
\end{equation} 
for the inverse map, 
where $r$ is chosen to be real, so 
the map $z(w)$ is unique. 
This map and the map to the complex conjugate domain 
$\bar D_{-}$, 
\begin{equation} 
\label{z2} 
\bar z(t,w^{-1})= r(t)\,w^{-1} + 
\sum_{k=0}^{\infty} \bar u_k(t)\,w^k\,, 
\end{equation} 
resolve the unitary condition for the Schwarz function 
and give it the following interpretation. 
If $w$ is the image of a point $z$, then 
$S(z)$ is the complex conjugate pre-image of $w^{-1}$: 
$S(z)=\bar z(w^{-1}(z))$. 
 
{\it 4.} The  idealized LG problem 
has an instructive form in terms of the Schwarz function: 
\begin{equation} 
\label{k} 
\partial_t S =\frac{q}{\pi }\partial_z\mbox{log}\,w. 
\end{equation} 
To derive (\ref{k}) (following \cite{H}), we 
differentiate $\bar z(t,w^{-1}) = S(t, z(t,w))$. We get 
$\partial_t \bar z(t,w^{-1}) = 
\partial_t S(t,z) + \partial_z S(t,z)\partial_t z(t,w)$ 
and, by virtue of (b), 
$V_n = {\rm Im}(\bar z_t z_l) = S_t/(2i\sqrt S_z )$. 
From $\log w (z)=-2\pi p\,(x,y)/q +i\phi\,(x,y)$ and $p=const$ along 
${\cal C}(t)$, we conclude that 
$-\partial_n p = qw_z/(2\pi iw \sqrt S_z)$. Since $V_n = -\partial_n p$, we 
obtain (\ref{k}). 
From now on we set $q=\pi$ by making a proper time rescaling. 
 
The equation (\ref{k}) written in terms 
of the conformal maps (\ref{z1}), (\ref{z2}) has the form \cite{1}: 
\begin{equation} 
\label{zz} 
\{z(t,w),\,\bar z(t,w^{-1})\}=1, 
\end{equation} 
where we define the Poisson bracket by $\{f,\,g\}\equiv 
w(\partial_w f \partial_t g \! -\! \partial_t f \partial _w g)$. 
On comparing 
powers of $w$ for both sides of (\ref{zz}), we get a set of equations for the 
coefficients $u_k, \bar u_k$ of the 
Laurent series, (\ref{z1}) and (\ref{z2}), with 
fixed $C_k$. Eq.\,(\ref{zz}) suggests 
the Hamiltonian structure of the problem: 
$z(t,w)$, $\bar z(t, w^{-1})$ and $\log w , \,t$ are canonical pairs.

{\it 5.} Consider the function $\Omega (z)$ defined on the curve by 
\begin{equation} 
\label{mj} 
\Omega(z) = \frac{|z|^2}{2} +2iA(z),\;\;\;z\in {\cal C}\,, 
\end{equation} 
where we have separated the real and imaginary parts. 
Here $A(z)$ is the 
area of the sector enclosed by ${\cal C}$ and 
bounded by the ray $\mbox{arg}\, z$ and some fixed reference ray. 
Just like the Schwarz function, this function can be analytically 
continued within a strip containing the curve. 
Indeed, writing (\ref{mj}) in terms of $S(z)$, we have 
\begin{eqnarray}  &&
\label{m} 
\Omega(z) \!=\! 
zS(z)/2 \!+\! 2i \! \int\limits^z \![S(z') dz' \!-\! z' dS(z')]/4i\!=\! 
\int\limits^z \!\!S(z')dz' 
\nonumber\\ &&
\!=\!\sum_{k=1}^\infty \!C_k z^k/\pi \!+\! t\log z \!-\! v_0/2 \!-\! 
\sum_{k=1}^\infty \! C_{-k}z^{-k}/\pi , 
\end{eqnarray} 
where $v_0$ does not depend on $z$. 
As is seen from the above definition, $\Omega$ is defined 
up to a purely imaginary $z$-independent term. We fix 
it by requiring $v_0$ to be real. 
 
The function $\Omega$ is the generating function of the 
canonical transformation 
$(\log w,\,t)\to (z,\,\bar z)\;$. Indeed, from (\ref{m}) 
we have 
$S(z)=\partial_z\Omega(z)$ and, by virtue of (\ref{k}), 
$\log w=\partial_t\Omega$. 
Therefore, the differential 
$d \Omega = S(z)\,dz + \log w \, dt$ 
encodes the LG equations (\ref{k}) or (\ref{zz}). 
 
{\it 6.} Now we extend the differential $d \Omega$ to include 
variations of all higher moments. For $k\geq 1$, let us denote 
\begin{equation} 
\label{H} 
t_k = \frac{C_k}{\pi k},\,\, 
H_k=\frac{\partial\Omega(z)}{\partial t_k},\,\, 
\bar H_k=- \frac{\partial\Omega(z)}{\partial \bar t_k}\,. 
\end{equation} 
Then the multi-time Hamiltonian system is defined as 
$$d\Omega = S(z)\,dz + \log w \, dt + \sum_{k=1}^{\infty} 
 (H_k dt_k - \bar H_k d \bar t_k)\,.$$ 
Thus, the flows with respect to the ``times'' $t_k$ are 
\begin{equation} 
\label{g} 
\partial_k z = \{ z, H_k \}\,, \;\;\;\; 
\partial_{\bar k} z =\{ z, \bar H_k \}\,, 
\end{equation} 
where $\partial_k = \partial/\partial t_k$, $\partial_{\bar k} = 
\partial/\partial \bar t_k$, and $\{, \}$ is the canonical Poisson bracket 
introduced above. These equations are consistent due the symmetry relations 
$\partial _l H_k(z) = 
\partial _k H_l(z)$ 
which follow from (\ref{H}). In terms of $w$, these conditions have 
the form of the zero-curvature equations: 
\begin{equation} 
\label{zero} 
\partial _k\, H_l(w)- 
\partial _l\,H_k(w) = 
\{H_k\,(w), H_l\,(w)\} \,. 
\end{equation} 
We now proceed to calculate the Hamiltonians. 
Below we will prove 
that in addition to (\ref{H}), for $z$ on the curve, 
the Hamiltonians can be equivalently defined as 
\begin{equation} 
\label{1H} 
H_k=-\,\partial_k \bar \Omega(\bar z). 
\end{equation} 
(The derivative is taken at fixed $\bar z$.) 
Then (\ref{m}) gives us 
\begin{eqnarray} 
\label{p} 
H_k&=&z^k - \partial_k v_0/2 - 
\sum_{l=1}^{\infty}\partial_k v_l\, 
z^{-l}/l\, = \\ 
&=&\partial_k v_0/2 + 
\sum_{l=1}^{\infty} 
\partial_k \bar v_l\, 
\bar z^{-l}/l 
\label{p1}, 
\end{eqnarray} 
where we set $v_{k} = C_{-k}/\pi$. 
Eq.(\ref{p}) implies that the Laurent expansion 
of the $H_k$ at $w=0$ does 
not contain powers of $w$ higher than $w^k$. Moreover, all 
non-negative powers of $w$ come from the first two terms of 
Eq.(\ref{p}). In turn, Eqs.\,(\ref{p1}) and (\ref{z2}) imply that 
$H_k$ does not contain negative powers of $w$. 
Altogether they  mean that $H_k$  is a polynomial in 
$w$ of degree $k$. It reads 
\begin{equation} 
\label{n} 
H_k(w)= (z^k(w))_{+}+\frac{1}{2} (z^k(w))_0. 
\end{equation} 
The symbol $(f(w))_+$ means a truncated Laurent series, 
where only terms with positive powers of $w$ are kept; 
$(f(w))_0$ is the constant ($w^0$) part of the series. 
 
It remains to prove Eq.\,(\ref{1H}). We first notice that 
$\partial_{j}{\rm Re}\,\Omega (z)=0$ 
if $z$ belongs to the curve. This property 
is proved by 
differentiating the real part of (\ref{mj}), 
$\Omega(z) + \bar \Omega (\bar z) = |z|^2$. The analytic 
continuation away from the curve gives $\Omega (z)+ 
\bar \Omega (S(z))=zS(z)$. Taking the partial derivative 
with respect to $t_j$ and restricting the result to the curve 
again, we get: 
\begin{equation}\label{b} 
\partial_k \Omega(z) + 
\partial_k \bar\Omega(\bar z)+ \partial_k S(z)\,\partial_{\bar z} \bar\Omega 
(\bar z) = z \partial_k S(z). 
\end{equation} 
But the r.h.s. 
and the last term in the l.h.s. of (\ref{b}) are equal since 
$z =\bar S(\bar z) = 
\partial_{\bar z} \bar \Omega (\bar z)$. Thus (\ref{b}) reads 
\begin{equation} 
\label{imagin} 
\partial_k [\Omega(z) + \bar \Omega (\bar z)] =0, 
\end{equation} 
where $z$ belongs to the 
curve. Eq.\,(\ref{1H}) follows by virtue of (\ref{H}). 
Now we see that $H_k$ and $\bar H_k$ (defined 
as in (\ref{H})) 
are indeed complex conjugates on the curve. 
 
Eqs.\,(\ref{g}) or alternatively (\ref{zero}) together with (\ref{n}), 
(\ref{z1}), and (\ref{zz}) provide an algorithm generating equations for the 
coefficients of the conformal map (\ref{z1}). Two first Hamiltonians are $H_1 = 
rw + u_0/2$, and $H_2 = r^2w^2 + 2ru_0w + ru_1 + u_0^2/2$. The  first equation 
of the hierarchy is 
\begin{equation} 
\label{toda} 
\partial^{2}_{1 \bar 1} \varphi 
=\partial_t \exp (\partial_t \varphi ), 
\end{equation} 
where $r^2=\exp (\partial_t \varphi )$. One can see from $\partial_t \Omega = 
\log w$ that $\varphi$ is the constant term in the expansion 
(\ref{m}): $\varphi =v_0$. 
 
{\it 7.} 
The unitarity condition 
(a) for the Schwarz function 
and the properties (\ref{mj}), 
(\ref{imagin}) of the generating function $\Omega$ 
(which actually follow from the unitarity) 
impose important relations among
the harmonic moments of any smooth simply connected 
domain and the harmonic moments of its complement. 
First, from (\ref{mj}) one can derive the following 
sum rules: 
\begin{equation} 
\label{sr1} 
\sum_{k\geq 1}kt_kv_k= \! 
\sum_{k\geq 1}k\bar t_k\bar v_k\,, \;\;\; 
\bar v_1 \!=\! tt_1 \!+\! 
\sum_{k\geq 2}k t_k v_{k-1}\,. 
\end{equation} 
Second, there are symmetry 
relations for derivatives of the harmonic moments 
$v_k =C_{-k}/\pi $ 
of the interior domain $D_{+}$ with respect to the 
(rescaled) harmonic moments $t_j$ of the exterior domain: 
$\partial_j v_k=\partial_k v_j$, 
$\partial_{\bar j} v_k=\partial_k \bar v_j$, 
$\partial_t v_k=\partial_k v_0$. 
The proof: it follows from (\ref{p1}) 
that $\oint _{{\cal C}}H_j dH_k =0$ for 
all $j,k$, 
then it is easy to see that 
$$ 
\partial_j v_k= 
\oint _{{\cal C}} z^k dH_j/2\pi i 
=\oint _{{\cal C}} z^j dH_k/2\pi i= 
\partial_k v_j\,. 
$$ 
This implies that for each analytic curve ${\cal C}(t, t_j)$ there 
exists a real function (prepotential) $F(t, t_j, \bar t_j)$ such that 
\begin{equation} 
\label{F} 
v_j =\partial_j F \,,\;\;\;\; 
\bar v_j =\partial_{\bar j} F \,,\;\;\;\; 
v_0 =\partial_t F \,. 
\end{equation} 
This function determines $H_k(z)$ via (12). 

{\it 8.} Equations (\ref{g}), (\ref{n}) 
are familiar in the soliton 
literature as the dispersionless limit \cite{dToda} 
of the 2D Toda hierarchy. 
Dispersionless hierarchies of this kind \cite{Kri} 
are extensions of the 
integrable equations of hydrodynamic type \cite{hydro} 
to the multidimensional case. 
Many special solutions were found in \cite{sol}. 
Eq.\,(\ref{zz}) is known as the string 
equation \cite{string}. (See also 
\cite{AWM} for more recent developments 
in the 2D Toda hierarchy.) 
To make the contact we now review, 
following \cite{dToda}, the 
standard setup of the 2D Toda hierarchy 
and its dispersionless limit. 
 
The 2D Toda hierarchy is usually introduced 
by means of two difference Lax operators: 
\begin{eqnarray} 
\label{L} 
L &=& r(t) e^{\hbar \partial_t} + 
  \!\displaystyle{\sum_{k=1}^{\infty} u_k(t)} 
e^{-k\hbar \partial_t}, 
\nonumber\\ 
\bar L &=& r(t\!-\! \hbar ) e^{-\hbar \partial_t} + 
  \!\displaystyle{\sum_{k=1}^{\infty}} \bar u_k(t) 
e^{-k\hbar \partial_t}, 
\end{eqnarray} 
where $r,\,u_k$ and $\bar u_k$ are functions of $t$ 
and of two sets of independent parameters 
$t_k,\, \bar t_k,\;k>0$. 
These functions obey the Lax-Sato equations: 
\begin{eqnarray}\label{S} 
&&\hbar\partial_{t_k} L = [L,H_k], \;\;\; {\rm where}\,\,H_k = 
(L^k)_+ \!+\! (L^k)_0/2, 
\nonumber\\ 
&&\hbar\partial_{\bar t_k} \bar L 
= [\bar H_k,L], \;\;\; {\rm where}\,\,{\bar H_k} 
= (\bar L^{k})_- \!+\! (\bar L^k)_0/2 \,. 
\end{eqnarray} 
The symbol $ (L^k)_{\pm}$ means the part of the operator that 
consists of positive (negative) powers of the shift operator 
$e^{\hbar \partial_t}$, 
and  $ (L^k)_0$ is the part 
that does not contain the shift operator. 
The first equation of the hierarchy is the familiar 
2D Toda equation: 
\begin{equation} 
\label{T} 
\partial^{2}_{1 \bar 1}\varphi (t) 
=e^{\varphi (t+\hbar )-\varphi(t)}-e^{\varphi (t)- 
\varphi(t-\hbar)}\,, 
\end{equation} 
where 
$r^2 (t)=e^{\varphi (t+\hbar ) -\varphi (t)}$. 
It is also customary to consider the Orlov-Shulman 
operators \cite{Or} 
\begin{eqnarray} 
&&M = \displaystyle{ 
\sum_{k=1}^{\infty}kt_kL^k + t +\! \sum_{k=2}^{\infty}} v_k L^{-k}, 
\nonumber\\ 
&&\bar M = \displaystyle{ 
\sum_{k=1}^{\infty}k\bar t_k\bar 
L^k + t +\! \sum_{k=2}^{\infty}} \bar v_k \bar L^{-k}, 
\end{eqnarray} 
where $v_k=v_k(t, t_j , \bar t_j)$, 
$\bar v_k =\bar v_k (t, t_j , \bar t_j)$ are functions 
such that the operators obey the conditions 
$[L,M]=\hbar L, \,\,\,\,[\bar L, \bar M] = -\hbar\bar L$. 
These operators satisfy the following linear equations: 
$L\Psi=z\Psi$, $\partial_k \Psi=H_k \Psi$, 
$\hbar z\partial_z \Psi= M\Psi$, 
and similarly for the bar-operators acting on $\bar \Psi$. 
 
{\it 9.} One particular solution of the 2D 
Toda hierarchy describes 2-matrix models. 
Consider the integral over two hermitian $N\times 
N$ matrices 
\begin{equation} 
\label{M} 
\tau=\int e^{Ntr(M\bar 
M+\sum_{k>0} (t_kM^k+\bar t_k \bar M^k))}dMd\bar M 
\end{equation} 
(the  partition function). 
It has been shown that this integral is the $\tau$-function for a 
special solution of the 2D Toda hierarchy with $\hbar=1/N$ 
\cite{GMMMO}. 
The solution is selected by the string equation 
$[L,\bar L]=\hbar$. 
The coefficients $v_k$ of 
the $M$-operator are given by 
$v_k=\partial_k \log\tau=<M^k>$, where 
$<\cdots>$ means an average over matrices with the weight 
(\ref{M}). A scaling behavior of a proper large $N$ limit of the 
matrix model is expected to 
describe 2D gravity \cite{GM}. 
 
The dispersionless hierarchy is 
obtained in the limit $\hbar\to 0$. In this limit, 
the shift operator $e^{\hbar \partial_t}$ is 
replaced by a classical variable $w$, the Lax operators 
are substituted by their 
eigenvalues $L\to z(w)$ and $\bar L\to \bar z(w^{-1})$, and the 
operator, $L^{-1}M$, becomes a function $S(z)$. 
At the same time all 
commutators are replaced by the Poisson brackets with the 
symplectic structure $\{w,t\}=w$, so 
$[L,\bar L]=\hbar$ turns into (\ref{zz}). 
Eq.\,(\ref{toda}) is the $\hbar\to 0$ limit of Toda's 
equation (\ref{T}). 
The wave function $\Psi$ is 
replaced by $e^{\Omega/\hbar}$, where $\Omega$ is the generating 
function of the canonical transformation 
$(\log w , t)\to (z, \bar z)$. 
At last, the $\tau$-function is 
$\tau =e^{F/\hbar^2}$ as $\hbar \to 0$, where the function $F$ is the 
prepotential introduced in (\ref{F}). 
 
{\it 10.} To summarize, comparing the semiclassical limit of the Toda 
Eqs.(\ref{L}), (\ref{S}), and the string equation, with Eqs.(\ref{z1}), 
(\ref{z2}), (\ref{g}), (\ref{n}), (\ref{zz}) 
of an arbitrary interface dynamics, 
we find an exact equivalence between them. This is the main result of this 
work. 
The sum rules (\ref{sr1}) for the harmonic moments are nothing 
else but (a part of) the $W$-constraints for the $\tau$-function. It also 
follows from above that the interface dynamics is equivalent to the $N \to 
\infty$ planar limit of the matrix model (\ref{M}): the (logarithm of) 
the partition function of the latter is the prepotential 
function $F$ (\ref{F}). 
 
It is tempting to understand what do the robust scaling behavior observed 
in variety of growth problems \cite{2} and a scaling behavior of 2D 
gravity \cite{GM},\cite{B}, have in common. 
 
We acknowledge useful discussions with 
B. \-Dub\-ro\-vin, G. Doolen,
J. \-Gib\-bons, L. \-Ka\-da\-noff, 
V. \-Ka\-za\-kov, I.~Kri\-che\-ver, 
L. \-Le\-vi\-tov, S. \-No\-vi\-kov, A. \-Or\-lov, 
B. \-Shrai\-man and T. \-Ta\-ke\-be. P. W.  would like to thank the Lady 
Davis foundation for the hospitality in Hebrew University in 
Jerusalem, where this work has been completed. P. W. was supported by 
grants NSF DMR 9971332 and MRSEC NSF DMR 9808595. 
The work of A. Z. was partially supported RFBR grant 
98-01-00344.

\end{document}